\documentclass[aps,prd,a4paper,superscriptaddress,amsfonts,amssymb,amsmath]{revtex4}
\usepackage{amssymb,latexsym}
\usepackage{amsmath, amsthm}
\usepackage{amscd}
\usepackage{times}
\usepackage{epsfig}
\usepackage{psfrag}
\usepackage{graphicx}
\usepackage{overpic}

\begin{document}

\title[]{Distinguishing black holes from
naked singularities through their accretion disk properties}

\author{Pankaj S. Joshi} \email{psj@tifr.res.in} \affiliation{Tata 
Institute of Fundamental Research, Homi Bhabha Road, Colaba, Mumbai 400005, India}
\author{Daniele Malafarina} \email{daniele@fudan.edu.cn}
\affiliation{Department of Physics, Fudan University, 220 Handan Road, 
Shanghai 200433, China}
\affiliation{Tata Institute of Fundamental Research, Homi Bhabha Road, 
Colaba, Mumbai 400005, India}
\author{Ramesh Narayan} \email{rnarayan@cfa.harvard.edu}
\affiliation{Harvard-Smithsonian Center for Astrophysics, 60 Garden
Street, Cambridge, MA 02138, USA}
\swapnumbers

\begin{abstract}

We show that, in principle, a slowly evolving gravitationally
collapsing perfect fluid cloud can asymptotically settle to a static
spherically symmetric equilibrium configuration with a naked
singularity at the center. We consider one such asymptotic final
configuration with a finite outer radius, and construct a toy model in
which it is matched to a Schwarzschild exterior geometry.  We examine
the properties of circular orbits in this model. We then investigate
observational signatures of a thermal accretion disk in this
spacetime, comparing them with the signatures expected for a disk
around a black hole of the same mass. Several notable differences
emerge.  A disk around the naked singularity is much more
luminous than one around an equivalent black hole. Also, the disk
around the naked singularity has a spectrum with a high frequency
power law segment that carries a major fraction of the total
luminosity.  Thus, at least some naked singularities can, in
principle, be distinguished observationally from black holes of 
the same mass. We discuss possible implications of these results.
\end{abstract}

\maketitle
\section{Introduction}
The no hair theorems in black hole physics state that a black hole is
completely characterized by three basic parameters, namely, its mass,
angular momentum, and charge.  Because of this extreme simplicity, the
observational properties of a black hole are determined uniquely by
these three intrinsic parameters plus a few details of the surrounding
environment.

There is compelling observational evidence that many compact objects
exist in the universe, and there are indications that some of these
might have horizons \cite{horizonevidence}. However, there is as yet
no direct proof that any of these objects is necessarily a black hole.
The dynamical equations describing collapse in general relativity do
not imply that the final endstate of gravitational collapse of a
massive matter cloud has to be a black hole (see \cite{review}); other
possibilities are also allowed. In the case of a continual collapse,
general relativity predicts that a spacetime singularity must form as
the endstate of collapse. However, recent collapse studies show
that, depending on the initial conditions from which the collapse
evolves, trapped surfaces form either early or late during the
collapse. Correspondingly, the final singularity may either be
covered, giving a black hole, or may be visible as a naked
singularity. If astrophysical objects of the latter kind
hypothetically form in nature, it is important to be able to
distinguish them from their black hole counterparts through
observational signatures.

General relativity has never been tested in the very strong field
regime, and very little is known about how matter behaves towards the
end of the gravitational collapse of a massive star, when extremely
large densities are reached and quantum effects possibly become
relevant.  Over the years, it has often been suggested that some
exotic stable state of matter might occur below the neutron degeneracy
threshold, allowing for the existence of quark stars, or boson stars,
or even more exotic astrophysical objects (see for example
\cite{exotic}).

At present we cannot rule out the possibility that such compact
objects do exist. On the other hand, the densities and sizes of
compact objects in the universe vary enormously, depending on their
mass. While a stellar mass quark star could have a density greater
than that of a neutron star, a supermassive compact object at the
center of a galaxy might have a density comparable to that of ordinary
terrestrial matter. This means that it is difficult to come up with a
single paradigm for all compact objects by simply modifying the
equation of state of matter.  Also, the processes that lead to the
formation of stellar mass compact objects are different from those
leading to supermassive objects. For one thing, time scales are very
different. Stellar mass compact objects form in a matter of seconds,
when the core of a massive dying star implodes under its own
gravity. We know little about the physical processes that lead to the
formation of supermassive compact objects, but whatever it is, it
doubtless operates far more slowly than in the stellar-mass case.

In our view it is important to study viable theoretical models that,
under reasonable physical conditions, lead to the formation of
different kinds of compact objects, and to investigate the properties
of these different end states.  In recent times, much attention has
been devoted to the observational properties of spacetimes that
describe very compact objects or singularities where no horizon is
present \cite{Ruffini}.  Many valid solutions of Einstein field
equations exist which describe spacetimes that are not black
holes. These are either vacuum solutions with naked singularities or
collapse solutions in the presence of matter. Naked singularity models
include the Reissner-Nordstrom, Kerr and Kerr-Newman geometries with a
range of parameter values that differentiate the black hole and naked
singularity regimes. Collapse models include dust collapse, models
with perfect fluids and those with other equations of state
\cite{review}.

Note further that the classical event horizon structure of the Kerr
metric can be altered in many ways. One way is by overspinning the
Kerr black hole in order to obtain a naked singularity
\cite{Kerr}. Another is by introducing deformations or scalar fields
to alter the spacetime and thus expose a naked singularity
\cite{Harko}.  Although the physical viability of some of these
examples is not clear, the fact remains that classical general
relativity allows for the formation of naked singularities in a
variety of situations.  If singular solutions represent a breakdown of
the theory in the regime of strong gravity, then a study of
some of these models might provide clues to new physics.  For example,
string theory or quantum gravity corrections can remove the Kerr
singularity, leaving open the possibility of non singular
superspinning solutions \cite{Horava}.

In a recent paper \cite{JMN}, the present authors showed that
equilibrium configurations describing an extended compact object can
in principle be obtained from gravitational collapse.  The models we
described correspond to a slowly evolving collapsing cloud which
settles asymptotically to a static final configuration that is either
regular or has a naked singularity at the center.  In this context, a
naked singularity must be understood as a region of arbitrarily large
density that is approached as comoving time goes to arbitrarily large
values. We briefly examined the properties of accretion disks around
the naked singularity solutions. There were significant physical
differences compared to disks around black holes, and it followed that
there could be astrophysical signatures that could distinguish black
holes from naked singularities.

In the above previous work, we considered the case of purely
tangential pressure and vanishing radial pressure.  Are the solutions
we obtained a consequence of this extreme simplification, or are they
representative of a generic class of solutions that survive even for
more reasonable equations of state?  We answer this question here at
least partly by exploring gravitational collapse of perfect fluid
objects with an isotropic pressure. Even in this more realistic case,
we find that objects with either regular interiors or naked
singularities form readily as a result of gravitational collapse. We
explore the similarities and differences between these new solutions
and the earlier solutions which had purely tangential pressure.

Our main purpose in the present paper is to investigate 
whether the naked singularity
models derived here and in our previous work can be observationally
distinguished from black holes of the same mass. We therefore take a
further step in this direction by calculating the spectral energy
distributions of putative accretion disks. We show that important
differences exist in the physical properties of accretion disks
around naked singularities compared to those around black holes, which
may help us distinguish black holes from naked singularities through
observations of astrophysical objects.
Although the toy models considered here are unlikely to be realized physically, some general features of these objects are revealed by our analysis and show that naked singularities could be observationally distinguished from black holes.

The structure of the paper is as follows: In section II we describe a
procedure by which a static perfect fluid object with a Schwarzschild
exterior metric can be obtained via gravitational collapse from
regular initial conditions.  In section III, we use the above
procedure to obtain a toy model of a static final configuration with a
naked singularity at the center. We then describe the properties of
accretion disks in this toy spacetime and compare these models to
disks around a Schwarzschild black hole of the same mass.  We also
briefly discuss other density profiles of astrophysical interest that
could be studied within the same framework. In the final section IV,
we discuss possible applications to astrophysical observations.

\section{Gravitational collapse} \label{collapse}
Spherical collapse in general relativity can be described by a
dynamical spacetime metric of the form
\begin{equation}\label{eq1}
    ds^2=-e^{2\nu}dt^2+\frac{R'^2}{G}dr^2+R^2d\Omega^2 \; ,
\end{equation}
where $\nu$, $R$ and $G$ are functions of the comoving coordinates $t$
and $r$.  For the perfect fluid case, the energy-momentum tensor is
given by $T^0_0=\epsilon, \; T_1^1=T_2^2=T_3^3=p$. The Einstein
equations then take the form
\begin{eqnarray}\label{p}
  p &=&-\frac{\dot{F}}{R^2\dot{R}} \; , \\ \label{rho}
  \epsilon&=&\frac{F'}{R^2R'} \; , \\ \label{phi}
  \nu'&=&-\frac{p'}{\epsilon+p} \; ,\\ \label{Gdot}
  \dot{G}&=&2\frac{\nu'}{R'}\dot{R}G \; ,
\end{eqnarray}
where $(')$ denotes a derivative with respect to comoving radius $r$
and $(\dot{})$ denotes a derivative with respect to time $t$ 
in the $(r,t)$ representation.  The
function $F$, called the Misner-Sharp mass, describes the amount of
matter enclosed within the shell labeled by $r$ at the time $t$, and is given by
\begin{equation}\label{misner}
F=R(1-G+e^{-2\nu}\dot{R}^2) \; .
\end{equation}

Equations (\ref{p})--(\ref{misner}) give five relations among the six
unknown functions $p(r,t)$, $\epsilon(r,t)$, $\nu(r,t)$, $G(r,t)$,
$F(r,t)$ and $R(r,t)$. We thus have the freedom to specify one free
function.  An assumed equation of state relating pressure to energy
density during the evolution of the system would fix this remaining
freedom and give a closed set of equations. It could happen, however,
that this approach leads to an analytically intractable problem.
Moreover, we do not know if the collapsing matter will have the same
unchanged equation of state as it evolves to higher and higher
densities as the collapse progresses.  There is in fact no
astrophysical or mathematical requirement that the equation of state
must be fixed as the collapse evolves in time.  We therefore prefer
here instead to choose the functional form of $F(r,t)$, which
corresponds implicitly to fixing the equation of state, which could
however change with time and space coordinates.  We choose a scenario
such that we approximate some standard equation of state at early
times, switching to some other, possibly exotic, kind of matter at
later times.

We use the scaling degree of freedom in the definition of $R$ to fix
the initial condition $R(r,t_i)=r$, where $t_i$ is the initial
time. To describe collapse we require $\dot{R}<0$, which guarantees
that $R$ decreases monotonically with respect to $t$.  Hence, we may
change coordinates from $(r,t)$ to $(r,R)$, thus in effect considering
$t=t(r,R)$.  Correspondingly, we can view the original function
$F(r,t)$ as a function of $r$ and $R$ and write $F=F(r,R)$.  In the
following, we use $(,r)$ to denote a derivative with respect to $r$ in
the $(r,R)$ representation, i.e.,
\begin{equation}
F'=F_{,r}+F_{,R}R'. \label{primecomma}
\end{equation}

The total mass of the system is not conserved during collapse, unless
one requires the further condition that $F(r_b,t)={\rm
  const}$, where $r_b$ corresponds to the boundary of the system. 
Therefore, we cannot in general match a collapse solution
to an exterior Schwarzschild metric. However, matching to a
generalized Vaidya spacetime at the boundary $R_b(t)=R(r_b,t)$ is
always possible \cite{matching}.

The procedure to solve the above system of Einstein equations is the
following. We choose the free function $F(r,R)$ globally and use
equations \eqref{p} and \eqref{rho} to obtain $p(r,R(r,t))$ and
$\epsilon(r,R(r,t))$
as functions of $r$, $R$ and its derivatives. 
We then integrate equation \eqref{phi} to obtain
$\nu$,
  \begin{equation}\label{int-nu}
    \nu(r,R)=-\int_0^r\frac{p'}{\epsilon+p}d\tilde{r},
  \end{equation}
and integrate equation \eqref{Gdot} to obtain $G$,
  \begin{equation}\label{G}
    G(r,R)=b(r)e^{2\int_r^R\frac{\nu'}{R'}d\tilde{R}}.
  \end{equation}
The free function $b(r)$ results from the integral in equation
\eqref{G}; it is related to the velocity profile in the collapsing
cloud. The integral in equation \eqref{int-nu} again gives a free
function of $t$, but this can be absorbed via a redefinition of the
time coordinate.  Once we have $\nu(r,R)$ and $G(r,R)$
as functions of $r$, $R$ and its derivatives, we can
integrate the Misner-Sharp mass equation \eqref{misner}
that becomes a differential equation involving $R$ and its derivatives.
We can write it in the form
  \begin{equation}\label{motion}
    t_{,R}=-\frac{e^{-\nu}}{\sqrt{\frac{F}{R}+G-1}},
  \end{equation}
and its integration gives $t(r,R)$, or equivalently $R(r,t)$ thus solving the system.

Substituting equations \eqref{int-nu} and \eqref{G} in equation
\eqref{eq1}, the metric of the collapsing spacetime takes the form
\begin{equation}\label{metric}
    ds^2=-e^{-2\int_0^r\frac{p'}{\epsilon+p}d\tilde{r}}dt^2+
\frac{R'^2}{b(r)e^{2\int_r^R\frac{\nu'}{R'}d\tilde{R}}}dr^2+R^2d\Omega^2
    \; .
\end{equation}
It may not always be possible to fully integrate the system of
Einstein equations globally. However, this is not always needed,
because by considering the behaviour of the functions involved, it is
often possible to extract useful information about collapse and to
integrate the solution at least in a neighborhood close to the center.

Typically some restrictions are required in order for the collapse
model to be considered physically viable. They are:
\begin{enumerate}
  \item Absence of shell crossing singularities, which arise from a
    breakdown of the coordinate system at locations where collapsing
    shells intersect. This requirement implies $R'>0$. We note in this
    connection that there is always a neighborhood of the central line
    $r=0$ of the collapsing cloud which contains no shell-crossing
    singularities throughout the collapse evolution \cite{shell}.
 
 \item Energy conditions, of which the usual minimal requirement is
    the weak energy condition, viz., positivity of the energy density
    ($\epsilon>0$) and of the sum of density and pressure
    ($\epsilon+p>0$).  This requires $F'(r,R)>0$ and $F_{,r}(r,R)>0$.
  \item Regularity at the center during collapse
    before the formation of the singularity. This includes the
    requirements that forces and pressure gradients vanish at the
    center and that the energy density has no cusps at $r=0$. The corresponding
    requirements are $F(r,R)\simeq r^3m(r,R)$ near $r=0$, $m'(0,R)=0$
    and $p'(0,R)=0$.
  \item Absence of trapped surfaces at the initial time. This last
    requirement is given by positivity of
    $\left[1-(F/R)\right]_{t=t_i}$ and translates to
    $\dot{R}^2_{t=t_i}<\left(e^{2\nu}G\right)_{t=t_i}$.
\end{enumerate}

Our aim now is to construct a dynamical collapse evolution 
such that the pressure eventually balances gravitational attraction 
and the collapsing object settles into a static configuration. 
Such a scenario is of course not always possible in gravitational
collapse of a massive matter cloud, and there are matter
configurations that can only collapse indefinitely without achieveing 
any possible equilibrium. Such is the case of collapse of a
pressureless dust cloud 
\cite{NS}, 
or similar systems where pressure is
very insignificant. Another possibility in collapse is that the 
cloud bounces back after reaching a minimum radius.  Typically, 
since we begin with a collapsing configuration,
for each shell labeled by the comoving radius $r$, three 
different behaviours are possible:
\begin{enumerate}
  \item Collapse: $\dot{R} < 0$. If $\dot{R}$ is negative at all times the shell will
    collapses to a central singularity.
  \item Bounce: $\dot{R}=0$ at a certain time, with $\ddot{R} > 0$ at this time.
    The shell bounces back and re-expands.
  \item Equilibrium configuration: $\dot{R}=\ddot{R}=0$. Here the
    collapse slows down and the shell achieves a static configuration.
\end{enumerate}
Shells that achieve an equilibrium configuration in a certain sense 
mark the separation between the region that collapses indefinitely 
and the region that eventually bounces back
\cite{bounce}. 
Both collapse and bounce 
would occur typically very rapidly, on a dynamical time scale which 
is proportional to the mass. Even in the case of supermassive compact 
objects, this time is much shorter than other typical time scales.

It would appear that all three possibilities above represent generic
behavior during gravitational collapse, depending on the masses and
velocities of the collapsing shells and the physical scenario involved
in collapse. For example, for a very massive star, an indefinite
collapse would seem inevitable if the star cannot shed away enough of
its mass in a very short collapse time scale to achieve any possible
equilibrium. On the other hand, for a much larger mass scale such as
galactic or yet larger scales, the collapse could proceed much more
slowly and even arrive at an equilibrium to form a stable massive
object. It is well known of course that gravitational collapse plays a
key role in the formation of large scale structures in the
universe. In many such cases, an evolving collapse would slow down,
eventually to form a stable massive object. In such cases, scenarios
such as the one considered here could be relevant.

We note that in order to achieve an equilibrium object as the endstate
of collapse or in order to have a useful quasi-equilibrium
configuration, it is necessary that each collapsing shell in the cloud
must individually achieve the condition $\dot{R}=\ddot{R}=0$.  Such an
object can be well approximated by a static configuration.  The
equation of motion \eqref{misner} can be written in terms of $R$, for
any fixed comoving radius $r$, in the form of an effective potential:
\begin{equation}\label{V}
    V(r,R)=-\dot{R}^2=-e^{2\nu}\left(\frac{F}{R}+G-1\right) \; .
\end{equation}

We already know that no static configuration is possible for the
pressureless (dust) case, where $V$ is negative at all times.  When
there is pressure, it is still possible for $V$ to be negative at all
times, giving continued collapse. However, other possibilities are
also allowed
since at any given time $t$ each shell $r$ can be either collapsing, expanding or still
giving rise to a wide array of scenarios.
We shall consider here the simple but very common case where $V$, as a function of $R$ for fixed $r$, is a polynomial of second order in $R$.
We see that one generic possibility in this case
 is that $V$ has two distinct
zeroes. In this case, the shell will bounce at a finite radius and
re-expand. Another possibility is that the two zeroes of $V$ coincide 
(namely $V$ has one root of double multiplicity) 
corresponding to an extremum at $V=0$. In this case, the shell will
coast ever more slowly towards the radius corresponding to $V=0$ and
will halt without bouncing having reached such radius with zero velocity 
and zero acceleration as $t$ goes to infinity. 
A global static configuration for this form of the potential is then
achieved only if each shell satisfies the condition that it has
one double root. It is not difficult to see that the velocity of the
cloud during collapse is always non-zero and therefore such an
equilibrium configuration can be achieved only in a limiting 
sense, as the comoving time $t$ goes to infinity 
(for a more detailed discussion on the condition leading
to equilibrium see \cite{JMN}).
For more general forms of the potential the allowed regions of 
dynamics for the shell and its behaviour are decided by the multiplicity of the roots of $V$.

The condition that the spacetime evolves towards an equilibrium
configuration is thus
\begin{equation}\label{static}
    \dot{R}=\ddot{R}=0,
\end{equation}
which is equivalent to 
\begin{equation}\label{staticV}
V=V_{,R}=0.
\end{equation}
From equation \eqref{V} we find
\begin{equation}
    V_{,R}=e^{2\nu}\left(\frac{F}{R^2}-\frac{F_{,R}}{R}+G_{,R}\right)
    -2\nu_{,R}e^{2\nu}\left(\frac{F}{R}+G-1\right) \; .
\end{equation}

The system achieves an equilibrium configuration if the solution of
the equation of motion \eqref{motion}, given by $R(r, t)$, tends
asymptotically to an equilibrium solution $R_e(r)$ such that the
conditions given in equations \eqref{static} or \eqref{staticV} are
satisfied.  Therefore, in order to have
\begin{equation}
    R(r,t)\xrightarrow[t\rightarrow \infty]{} R_e(r),
\end{equation}
we must choose the free function $F(r,R)$ in the dynamical collapse
scenario such that the quantities $F$, $\nu$, $G$ tend to their
respective equilibrium limits:
\begin{eqnarray}
  F(r,R) &\rightarrow& F_e(r)=F(r, R_e(r)), \\
  \nu(r,R) &\rightarrow& \nu_e(r)=\nu(r,R_e(r)), \\
  G(r,R) &\rightarrow& G_e(r)=G(r, R_e(r)).
\end{eqnarray}
Imposing the conditions \eqref{staticV}, we thus obtain two equations
which fix the behaviour of $G$ and $G_{,R}$ at equilibrium:
\begin{eqnarray}\label{ve}
  G_{e}(r)&=&1-\frac{F_e}{R_e} \; , \\ \label{Ge} 
  (G_{,R})_e  &=&G_{,R}(r, R_e(r)) = \frac{F_e}{R_e^2}-\frac{(F_{,R})_e}{R_e} \; ,
\end{eqnarray}
where the velocity profile $b(r)$ in equation \eqref{G} has been
absorbed into $G_e(r)$.

Note that at equilibrium the area radius $R$ becomes a monotonic 
increasing function of $r$. Therefore if
we define the new radial coordinate $\rho$ at equilibrium as
\begin{equation}
    \rho \equiv R_e(r), \;
\end{equation}
we can rewrite the functions at equilibrium as
\begin{eqnarray}
  F_e(r)&=&\mathrm{F}(\rho), \\
  \nu_e(r)&=&\phi(\rho), \\
  G_e(r)&=&\mathrm{G}(\rho)=1-\frac{\mathrm{F}}{\rho}.\label{Ger}
\end{eqnarray}
Then, from equations \eqref{rho} and \eqref{phi}, two of the Einstein
equations for a static source become
\begin{eqnarray}
  \epsilon(\rho) &=& \frac{\mathrm{F}_{,\rho}}{\rho^2}, \\
  p_{,\rho} &=& -(\epsilon+p)\phi_{,\rho},
\end{eqnarray}
where $({,\rho})$ denotes a derivative with respect to the new
static radial coordinate $\rho$.  The second equation is the
well known Tolman-Oppenheimer-Volkoff (TOV) equation.  The third
static Einstein equation, namely
\begin{equation}
    p=\frac{2\phi_{,\rho}}{\rho}\mathrm{G}(\rho)-\frac{\mathrm{F}(\rho)}{\rho^3},
\end{equation}
is obtained from equation \eqref{p} by imposing the equilibrium
condition and making use of equation \eqref{Gdot} at equilibrium.  The
metric \eqref{metric} at equilibrium then becomes the familiar static
spherically symmetric spacetime,
\begin{equation}\label{metric-eq}
ds^2=-e^{2\phi}dt^2+\frac{d\rho^2}{\mathrm{G}}+\rho^2d\Omega^2,
\qquad \rho \leq \rho_b\equiv R_e(r_b).
\end{equation}
This interior metric is matched to a Schwarzschild vacuum 
exterior at the boundary $\rho_b=R_e(r_b)$.  By matching
$g_{\rho\rho}$ at the matching radius $\rho_b$ and
making use of equation \eqref{Ger}, we see that the total
gravitational mass $M_T$ of the interior is given by
\begin{equation}\label{MT}
1-\frac{2M_T}{\rho_b} = G(\rho_b) = 1-
\frac{F(\rho_b)}{\rho_b}.
\end{equation}

Note that, in principle, the interior metric in equation 
\eqref{metric-eq} need not be regular at the center, as the eventual 
singularity is achieved as the result of collapse from regular 
initial data.  A singularity at the center of the static final 
configuration is then interpreted as a region of arbitrarily high 
density that is achieved asymptotically as the comoving time 
$t$ goes to arbitrarily large values.
Therefore when considering static interiors with a singularity
we are in fact approximating a slowly evolving configuration, where
shells have typically very `small' velocities and where the central region 
can reach very high densities.

Typically, for a static perfect fluid source of the Schwarzschild
geometry, various conditions could be required in order to ensure
physical reasonability \cite{Lake}.  Some of these are: the matter
satisfies an energy condition, matching conditions with the exterior
Schwarzschild geometry, vanishing of the pressure at the boundary,
monotonic decrease of the energy density and pressure with increasing
radius, sound speed within the cloud should be smaller than the speed
of light.  If the energy conditions are satisfied during collapse,
they will be satisfied by the equilibrium configuration as well, i.e.,
the positivity of the energy density and sum of density and pressure
at the origin follows from the same condition during collapse.  Also,
requiring only the weak energy condition allows the possibility of
negative pressures either during the collapse phase or in the final
equilibrium configuration.

Since we have required here the static configuration to evolve from
regular initial data, we may omit the condition that the final state
must be necessarily regular at the center. This leaves open the
possibility that a central singularity might develop as the
equilibrium is reached, where the singularity has to be understood in
the sense explained above. Absence of trapped surfaces at the initial
time ensures that, if a singularity develops, it will not be covered
by a horizon
\cite{initial}. 
In fact it can be shown that if trapped surfaces do form
at a certain time as the cloud evolves, then the collapse cannot be
halted and the whole cloud must collapse to a black hole or to a
singularity where the first point of singularity is visible but the
later portions of the singularity become covered in a black hole. We note
that some of the above conditions, although desirable, may be
neglected in special cases.  Many interior solutions are available in
the literature describing a static sphere of perfect fluid matching
smoothly to a Schwarzschild exterior geometry
(see \cite{DL} for a list of solutions or \cite{visser} for an algorithm to construct
interior solutions). 
As discussed above, we
can construct dynamically evolving collapse scenarios that lead
asymptotically to the formation of such static configurations.  If we
are able to achieve a static configuration from collapse from regular
initial data, and if we view a singularity as a region where density
increases to arbitrarily high values, signaling a breakdown of the
ability of general relativity to model the spacetime, we may then
neglect the requirement of regularity at the center.

\section{A toy model of a static spherically symmetric perfect 
fluid interior}\label{toy} 
In the following, we look for static
interiors with a singularity at the origin. We wish to investigate the
properties of such solutions and to establish whether such
hypothetical naked singularity objects could in principle be
distinguished in terms of observational signatures from black hole
counter-parts of the same mass.

We start with the most general static spherically symmetric metric
written in the form given in equation \eqref{metric-eq}.  As we have
seen, the perfect fluid source Einstein equations give
\begin{eqnarray}
  \epsilon &=& \frac{\mathrm{F}_{,\rho}}{\rho^2},\label{epsilon} \\ p &=&
  \frac{2\phi_{,\rho}}{\rho}\left[1-\frac{\mathrm{F}(\rho)}{\rho}\right]-\frac{\mathrm{F}(\rho)}{\rho^3},
    \\ p_{,\rho} &=& -(\epsilon+p)\phi_{,\rho},
\end{eqnarray}
where we have absorbed the factor $4\pi$ into the definition of
$\epsilon$ and $p$, and defined $\mathrm{F}_{,\rho}=d\mathrm{F}/d\rho$.
The third (Tolman-Oppenheimer-Volkoff) equation, combined with the
second equation, gives
\begin{equation}\label{OV}
    p_{,\rho}=-(\epsilon+p)\frac{\left[p\rho^3+\mathrm{F}(\rho)\right]}
{2\rho\left[\rho-\mathrm{F}(\rho)\right]}.
\end{equation}
Since the above system of Einstein equations consists of three equations
with four unknowns, in order to close the system we can either specify an
equation of state that relates $p$ to $\epsilon$ or use the freedom to
choose arbitrarily one of the other functions
\cite{wald}.  As discussed earlier, 
we opt for the latter approach and specify the form of the mass profile
$\mathrm{F}(\rho)$, which describes the final mass distribution
obtained from collapse.  Once we specify $\mathrm{F}$, equation
\eqref{OV} reduces to a first-order ordinary differential equation, 
which we can solve.

We follow the same procedure that we used in the pure tangential pressure
case \cite{JMN} and obtain a solution of the form
\begin{equation}
    \frac{\mathrm{F}(\rho)}{\rho}=M_0={\rm const}, \qquad
    \rho \leq \rho_b.
\end{equation}
By equation \eqref{MT}, this solution corresponds to a total mass
$M_T$ given by
\begin{equation}\label{MT2}
2M_T = \mathrm{F}(\rho_b) = M_0 \rho_b.
\end{equation}
Hence, to avoid a horizon, we require $M_0<1$.  Solving the first
Einstein equation \eqref{epsilon}, the energy density becomes
\begin{equation}
    \epsilon=\frac{M_0}{\rho^2}, \label{epsilon2}
\end{equation}
which clearly diverges as $\rho\to0$, indicating the presence of
a strong curvature singularity at the center.

The TOV equation \eqref{OV} now becomes
\begin{equation}
    p_{,\rho}=-\frac{(\epsilon+p)^2\rho}{2(1-M_0)}=
-\frac{(M_0+p\rho^2)^2}{2\rho^3(1-M_0)}, \label{pp}
\end{equation}
which can be integrated by defining the auxiliary parameter
\begin{equation}
    \lambda=\sqrt{\frac{1-2M_0}{1-M_0}}, \qquad
M_0 = \frac{1-\lambda^2}{2-\lambda^2}.
\end{equation}
Clearly, we require $\lambda\in[0,1)$, or equivalently $M_0<1/2$ (for
values of $M_0>1/2$ similar considerations apply with
$\lambda=\sqrt{-(1-2M_0)/(1-M_0)}$ but we do not consider this case
here).  In terms of $\lambda$, the
solution of equation \eqref{pp} can be written as
\begin{equation}
    p =
    \frac{1}{2-\lambda^2}\frac{1}{\rho^2}
\left[\frac{(1-\lambda)^2A-(1+\lambda)^2B\rho^{2\lambda}}{A-B\rho^{2\lambda}}\right],
\end{equation}
where $A$ and $B$ are arbitrary integration constants. Then, solving
the remaining Einstein equation gives
\begin{eqnarray}
  e^{2\phi} = (A\rho^{1-\lambda}-B\rho^{1+\lambda})^2.
\end{eqnarray}
This solution, among other similar interior static solutions, was
first investigated by Tolman in 1939 for a specific choice of
$\lambda$ 
\cite{Tolman}.  
The energy density of the solution may be rewritten as
\begin{equation}
    \epsilon = \left(\frac{1-\lambda^2}{2-\lambda^2}\right)
    \frac{1}{\rho^2}.
\end{equation}
The existence of a strong curvature singularity at the center can be
confirmed by an analysis of the Kretschmann scalar $K$ near
$\rho=0$, which gives
\begin{equation}
    K=\frac{16Q^4(1-\lambda^2)^2+\rho^4(Q_{,\rho}^2-2QQ_{,\rho\rho})^2
+8\rho^2Q^2Q_{,\rho}^2}{4\rho^4Q^4(2-\lambda^2)^2}, \qquad Q(\rho)=e^{2\phi(\rho)},
\end{equation}
which is clearly singular in the limit $\rho\to0$.

The static metric of the above solution takes the form
\begin{equation}\label{metric-sol}
    ds^2=-(A\rho^{1-\lambda}-B\rho^{1+\lambda})^2dt^2+(2-\lambda^2)d\rho^2
+\rho^2d\Omega^2, \qquad \rho \leq \rho_b. 
\end{equation}
It is matched at the boundary $\rho_b$ to a vacuum Schwarzschild
spacetime with total mass $M_{T}$.  Since $g_{\rho\rho}$ in the interior is
a constant, the matching will not in general be smooth, though it will
be continuous.  From the matching conditions for $g_{\rho\rho}$ and $g_{tt}$
at the boundary we obtain,
\begin{eqnarray}
  \rho_b &=& \frac{2(2-\lambda^2)}{(1-\lambda^2)} M_T, \label{rhob}\\ \label{B}
  B\,\rho_b^{2\lambda} &=& A-\frac{\rho_b^{\lambda-1}}{\sqrt{2-\lambda^2}}.
\end{eqnarray}
For all values of $\lambda \in [0,1)$, the pressure is maximum at the
  center and decreases outwards, becoming zero at a finite radius
  $\rho_p$ given by
\begin{equation}\label{boundary}
    \rho_p^{2\lambda}=\frac{A}{B}\frac{(1-\lambda)^2}{(1+\lambda)^2}.
\end{equation}
Since we require the pressure to vanish at the boundary of the cloud,
we obtain the further condition $\rho_b=\rho_p$, which,
together with equation \eqref{B}, fixes the two integration constants
$A$ and $B$:
\begin{eqnarray}
    A&=& \frac{(1+\lambda)^2\rho_b^{\lambda-1}}{4\lambda\sqrt{2-\lambda^2}}, \label{Acoeff} \\
    B&=& \frac{(1-\lambda)^2\rho_b^{-\lambda-1}}{4\lambda\sqrt{2-\lambda^2}}. \label{Bcoeff}
\end{eqnarray}

The sound speed inside the cloud is given by $c_s^2=\partial
p/\partial \epsilon$. We find
\begin{equation}
    c_s^2=\frac{p}{\epsilon}+\frac{4\lambda^2ABr^{2\lambda}}{(1-\lambda^2)(A-Br^{2\lambda})^2}, \qquad
\frac{p}{\epsilon}=\frac{A\frac{1-\lambda}{1+\lambda}-B\frac{1+\lambda}{1-\lambda}\rho^{2\lambda}}{A-B\rho^{2\lambda}},
\end{equation}
from which we see that $c_s(0)=(1-\lambda)(1+\lambda)<1$, as required,
and $c_s$ decreases as the radius $\rho$ goes from $0$ to $\rho_b$.
We may also rewrite the above relations in the form of an equation of
state, $p=p(\epsilon)$,
\begin{equation}
    p(\epsilon)=\epsilon\frac{A\left(\frac{1-\lambda}{1+\lambda}\right)\left(\frac{2-\lambda^2}
    {1-\lambda^2}\right)^\lambda \epsilon^\lambda-B\left(\frac{1+\lambda}{1-\lambda}\right)}
    {A\left(\frac{2-\lambda^2}{1-\lambda^2}\right)^\lambda \epsilon^\lambda-B}.
\end{equation}

In the limit $\lambda\to0$, corresponding to $M_0\to1/2$, we obtain
the most compact member of the above family of solutions. It has an
equation of state $p\approx\epsilon$ and hence sound speed $c_s\to1$.
Note, however, that at the outer boundary, $p\to0$ but $\epsilon$ does not
vanish, so the equation of state is not strictly isothermal.  For this
solution, the matching radius with the Schwarzschild exterior is at
$\rho_b=4M_T$, i.e., twice the Schwarzschild radius.  In the
case of pure tangential pressure, which we considered in our previous
paper \cite{JMN}, we found physically meaningful solutions down to
$\rho_b=3M_T$, and even more compact solutions with exotic
properties.  This is one respect in which the perfect fluid model
differs from the tangential pressure case.

For $\lambda=1/2$, corresponding to $M_0=3/7$, we obtain an equation
of state
\begin{equation}
    p=\frac{\epsilon}{3}\left(1-\frac{8B}{A\sqrt{\frac{7}{3}\epsilon}-B}\right),
\end{equation}
which approaches the radiation equation of state $p=\epsilon/3$ as
$\rho\rightarrow 0$. For this model, $\rho_b=(14/3)M_T$, i.e., the
object is a little less compact than the model with $\lambda\to0$.

Finally, the case $\lambda\to1$, $M_0\to0$, corresponds to
$\rho_b/M_T\to\infty$, and hence an infinitely large object.  In
this limit, our solution reduces to the classical Newtonian singular
isothermal sphere solution. Correspondingly, $A\to1$, $B\to0$, and the
metric in equation \eqref{metric-sol} reduces to the metric of flat
space.

Considering next the energy conditions, it is easy to see that $M_0>0$
implies positivity of the energy density and pressure. The pressure
decreases from its maximum value at the center to zero at the
boundary. From
\begin{equation}
    \epsilon+p=\frac{(1-\lambda)(1+\lambda)}{2-\lambda^2}
\frac{1}{\rho^2}\left(1+\frac{p}{\epsilon}\right),
\end{equation}
we see that the weak energy condition is satisfied throughout the
interior of the cloud.

\subsection{Properties of circular orbits}
We wish to investigate basic observational properties of accretion
disks orbiting in the above family of spacetimes. To this end, we
study circular geodesics. Our assumption is that a test particle
orbiting inside the ``cloud'' does not interact with the material of
the cloud but merely feels its gravitational influence
\footnote{Of course as the particles approach the center friction and viscosity 
will play an increasingly important
role thus deviating the behaviour from that of a classical accretion 
disk
\cite{gregoris}. Nevertheless we can 
assume that until a certain radius the approximations made here 
are satisfied.}.

Our perfect fluid model has five parameters, $M_T$, $\lambda$,
$\rho_b$, $A$, $B$, and there are three matching conditions at the
boundary $\rho=\rho_b$, viz., matching of $g_{tt}$ and $g_{\rho\rho}$
with the exterior Schwarzschild metric and the condition
$p(\rho_b)=0$.  Thus, we are free to choose two of the five
parameters. For convenience, in the following we choose the total
gravitational mass of the cloud $M_T$ to be equal to unity. This still
leaves one free parameter, which we choose to be $\lambda$. Once we
pick a value of $\lambda$, the boundary radius $\rho_b$ is given by
equation (\ref{rhob}), and the coefficients $A$ and $B$ are given by
equations (\ref{Acoeff}) and (\ref{Bcoeff}).

Given the spherical symmetry of the metric we can always choose the
coordinate $\theta$ such that the geodesic under consideration lies in
the equatorial plane ($\theta=\pi/2$). Time-like geodesics in this
plane satisfy
\begin{equation}
    -H(\rho)\left(\frac{dt}{d\tau}\right)^2+(2-\lambda^2)\left(\frac{d\rho}{d\tau}\right)^2
    +\rho^2\left(\frac{d\varphi}{d\tau}\right)^2=-1,
\end{equation}
where $H(\rho)=(A\rho^{1-\lambda}-B\rho^{1+\lambda})^2$.

From the two killing vectors, $\zeta^b$, $\eta^a$, associated with
time-translational symmetry and rotational symmetry, we can calculate
two conserved quantities, the energy per unit mass,
$E=g_{ab}\zeta^au^b=e^{2\phi}dt/d\tau=Hdt/d\tau $, and the angular momentum per
unit mass, $L=g_{ab}\eta^au^b=\rho^2d\varphi/d\tau$.
For circular geodesics we must have $d\rho/d\tau=0$. Therefore, we
obtain
\begin{eqnarray}
  E^2 &=& \frac{2H^2}{2H-H_{,\rho}\rho} , \\
  \frac{L^2}{\rho^2} &=&  \frac{H_{,\rho}\rho}{2H-H_{,\rho}\rho}.
\end{eqnarray}

\begin{figure}
\includegraphics[scale=0.9]{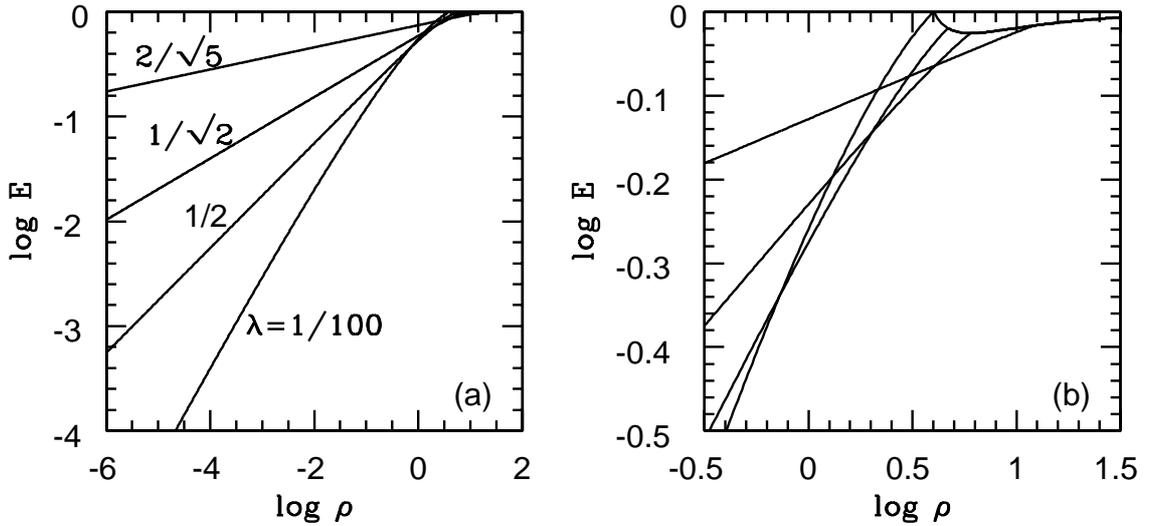}
\caption{(a) Energy per unit mass $E$ of circular orbits as a function
  of radius $\rho$ for four perfect fluid models of unit mass
  ($M_T=1$). The parameter $\lambda$ and the radius of the boundary
  $\rho_b$ of the four models are, respectively: $(\lambda,\ \rho_b)=
  (1/100,\ 4.0002)$; $(1/2,\ 14/3)$; $(1/\sqrt{2},\ 6)$;
  $(2/\sqrt{5},\ 12)$. The curves are labeled by their values of
  $\lambda$.  Each solution is matched to a Schwarzschild exterior at
  $\rho=\rho_b$. (b) Close up of the matching region between the
  perfect fluid interior and the Schwarzschild exterior. The models
  with $\lambda=1/100$, 1/2, have $\rho_b$ lying inside the innermost
  stable circular orbit $\rho_{\rm ISCO}=6$ of the Schwarzschild
  spacetime.  For these two solutions, $E$ decreases between $\rho_b$
  and $\rho_{\rm ISCO}$, indicating a zone of unstable circular
  orbits.}
\label{fig1}
\end{figure}

Figure \ref{fig1} shows the variation of the energy per unit mass $E$
with radius $\rho$ for a selection of models corresponding to
$\lambda=1/100$, $1/2$, $1/\sqrt{2}$, $2/\sqrt{5}$. The boundaries of
the four models are at (see eq.~\ref{rhob}) $\rho_b=4.0002$, $14/3$, 6
and 12. In all the models, $E$ goes to zero as $\rho\to0$, with a
power-law dependence: $E\sim \rho^{1-\lambda}$.  At $\rho=\rho_b$,
each model is matched to an exterior Schwarzschild spacetime with mass
$M_T=1$. Panel (b) in Figure \ref{fig1} shows the matching region.

By construction, $E$ is continuous across the matching boundary, but
$dE/d\rho$ is not. Note in particular that the vacuum Schwarzschild
spacetime has an innermost stable circular orbit (ISCO) at $\rho_{\rm
  ISCO}=6$. For $\rho<\rho_{\rm ISCO}$, $E$ increases with decreasing
$\rho$, which is one of the consequences of the absence of stable
orbits. The model with $\lambda=2/\sqrt{5}$ has its boundary at
$\rho_b=12$ and hence has stable circular orbits all the way from
large radii down to $\rho\to0$. So too does the model with
$\lambda=1/\sqrt{2}$ which has its matching radius at the ISCO,
$\rho_b=\rho_{\rm ISCO}$. However, the other two models
($\lambda=1/2$, $1/100$) have $\rho_b<\rho_{\rm ISCO}$. Hence, these
models have stable circular orbits inside $\rho_b$ and outside
$\rho_{\rm ISCO}=6$, but no stable orbits in between.

\begin{figure}
\includegraphics[scale=0.9]{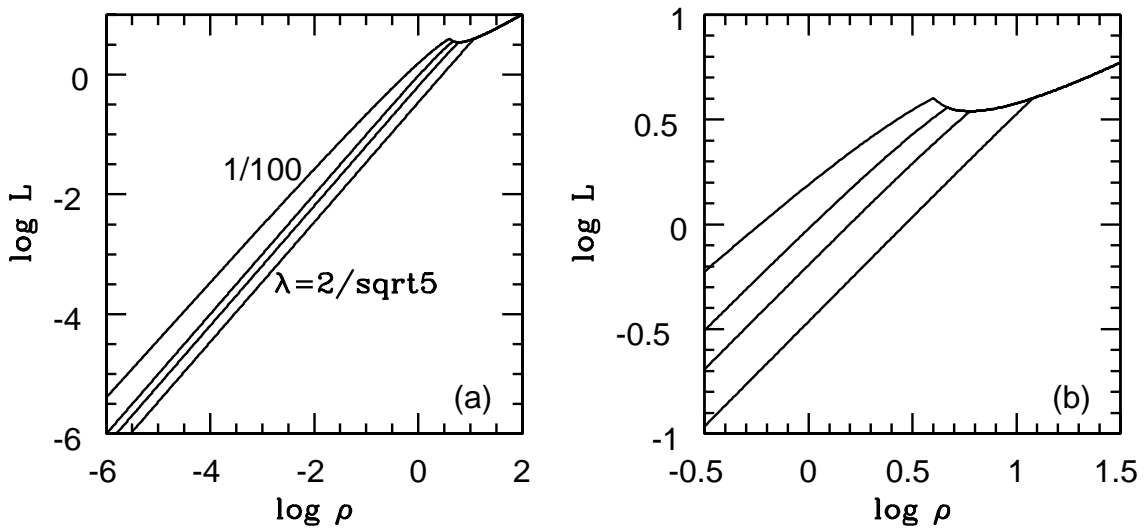}
\caption{(a) Angular momentum per unit mass $L$ of circular orbits as
  a function of radius $\rho$ for the same four perfect fluid models
  shown in Figure \ref{fig1}. The values of $\lambda$ are: $1/100$,
  $1/2$, $1/\sqrt{2}$, $2/\sqrt{5}$ (the curves corresponding to the
  first and last solutions are labeled). (b) Close up of the matching
  region between the perfect fluid interior and the Schwarzschild
  exterior. Note, as in Figure \ref{fig1}, that the solutions with
  $\lambda=1/100$, $1/2$ have $L$ decreasing between $\rho_b$ and
  $\rho_{\rm ISCO}=6$. Circular orbits are unstable over this range of
  radius.}
\label{fig2}
\end{figure}

Figure \ref{fig2} shows analogous results for the angular momentum per
unit mass $L$. Here, at small radii, $L\sim\rho$ for all the models.
As in the case of $E$, the angular momentum matches continuously
across the boundary $\rho=\rho_b$ to the external Schwarzschild
solution. As panel (b) shows, the two models with $\lambda=1/100$ and
$1/2$ have $dL/d\rho<0$ for a range of radii between $\rho_b$ and
$\rho_{\rm ISCO}$. This corresponds to the region with unstable
circular orbits.

\begin{figure}
\includegraphics[scale=0.9]{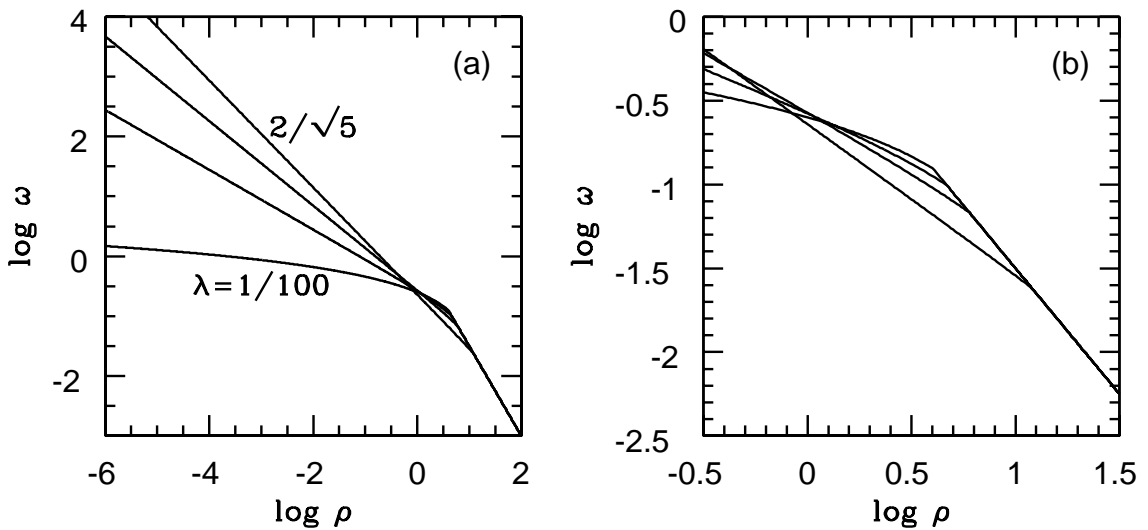}
\caption{(a) Angular velocity $\omega$ of circular orbits as a
  function of radius $\rho$ for the same four perfect fluid models
  shown in Figure \ref{fig1}. The values of $\lambda$ are: $1/100$,
  $1/2$, $1/\sqrt{2}$, $2/\sqrt{5}$ (the curves corresponding to the
  first and last solutions are labeled). (b) Close up of the matching
  region between the perfect fluid interior and the Schwarzschild
  exterior.}
\label{fig3}
\end{figure}

The angular velocity $\omega=d\phi/dt$ of particles on circular orbits
is given by
\begin{equation}
    \omega^2=-\frac{g_{tt,\rho}}{g_{\phi\phi,\rho}}=\frac{H_{,\rho}}{2\rho}.
\end{equation}
This quantity scales as $\omega\sim\rho^{-\lambda}$ as $\rho\to0$.
Figure \ref{fig3} shows a plot of $\omega$ vs $\rho$ for the same four
models as before.

\subsection{Properties of accretion disks}
Gas in an accretion disk loses angular momentum as a result of
viscosity and moves steadily inwards along a sequence of nearly
circular orbits. Using just the properties of circular geodesics, and
without needing to know the detailed properties of the viscous stress,
it is possible to calculate the radiative flux emitted at each radius
in the disk \cite{nt73}.

Since in our model both $E$ and $L$ tend to zero as $\rho$ goes to
zero, no energy or angular momentum is added to the central
singularity by the gas in the accretion disk.  The central singularity
may be considered to be `stable' in this sense.  Indeed, since
$E\to0$, all the mass energy of the accreting gas is converted to
radiation and returned to infinity, i.e., the net radiative luminosity
as measured at infinity satisfies ${\cal L}_\infty = \dot{m}c^2$,
where $\dot{m}$ is the rate at which rest mass is accreted. Accretion
disks around our model naked singularity solutions are thus perfect
engines that convert mass into energy with 100\% efficiency.

From the behaviour of $E$ and $L$ of circular geodesics
(Figs.~\ref{fig1} and \ref{fig2}), we can distinguish two different
regimes of accretion, depending on the value of $\lambda$:
\begin{itemize}
\item For $\lambda\in[1/\sqrt{2},1)$, corresponding to
  $\rho_b\in[6M_{T},+\infty)$, particles in the accretion disk follow
    circular geodesics of the Schwarzschild exterior until they
    reach the matching radius $\rho_b$ at the outer edge of the
    cloud. Inside the cloud, the particles switch smoothly and
    continuously to the circular geodesics of the interior
    solution. Thus, the accretion disk extends without any break from
    arbitrarily large radii down to the singularity $\rho\to0$.
\item For $\lambda\in(0,1/\sqrt{2})$, corresponding to
  $\rho_b\in(4M_{T},6M_{T})$, particles reach the ISCO of the exterior
  Schwarzschild spacetime at $\rho_{\rm ISCO}=6$. Inside this radius
  no stable circular orbits are allowed, so the gas in the disk
  plunges with constant $E=E_{\rm ISCO}$ and $L=L_{\rm ISCO}$ until it
  crosses the boundary of the cloud at $\rho_b$. Inside the cloud,
  circular geodesics are allowed again. The gas penetrates into the
  cloud until it reaches a radius $\rho_L$ at which the specific
  angular momentum of a local circular geodesic $L$ is equal to
  $L_{\rm ISCO}$. At this radius, the gas settles into a stable
  circular orbit and radiates away any excess energy. Further
  evolution than proceeds in the standard fashion, with the gas
  steadily moving to smaller radii until $\rho\to0$. The accretion
  disk is thus divided into two parts, one in the vacuum exterior over
  radii $\rho\ge\rho_{\rm ISCO}$ and the other inside the cloud over
  radii $\rho\le\rho_L$.
\end{itemize}

In the following, we focus on models belonging to the first regime,
where we have a continuous disk extending with no break from large
radius down to the singularity. Specifically, we consider models with
$\lambda=1/\sqrt{2}$, $2/\sqrt{5}$, which have matching boundaries at
$\rho_b=6$, $12$, respectively. The radiative properties of accretion
disks in these spacetimes may be calculated using the relations given
in \cite{nt73}. In the local frame of the accreting fluid, the
radiative flux emitted by the disk (which is the energy per unit 
area per unit time) is given by
\begin{equation}\label{flux}
    \mathcal{F}(\rho)=-\frac{\dot{m}}{4\pi \sqrt{-g}}\frac{\omega_{,\rho}}
{(E-\omega L)^2}\int^\rho_{\rho_{in}}(E-\omega L)L_{,\mathrm{\tilde{\rho}}}d\mathrm{\tilde{\rho}} \; ,
\end{equation}
where $\dot{m}$ is the rest mass accretion rate, assumed to be
constant, $\rho_{\rm in}$ is the radius of the inner edge of the
accretion disk, which is zero for our singular cloud models, and $g$
is the determinant of the metric of the three-sub-space
$(t,\rho,\phi)$,
\begin{equation}
g(\rho) = -(2-\lambda^2) \rho^2 H(\rho).
\end{equation}

\begin{figure}
\includegraphics[scale=0.9]{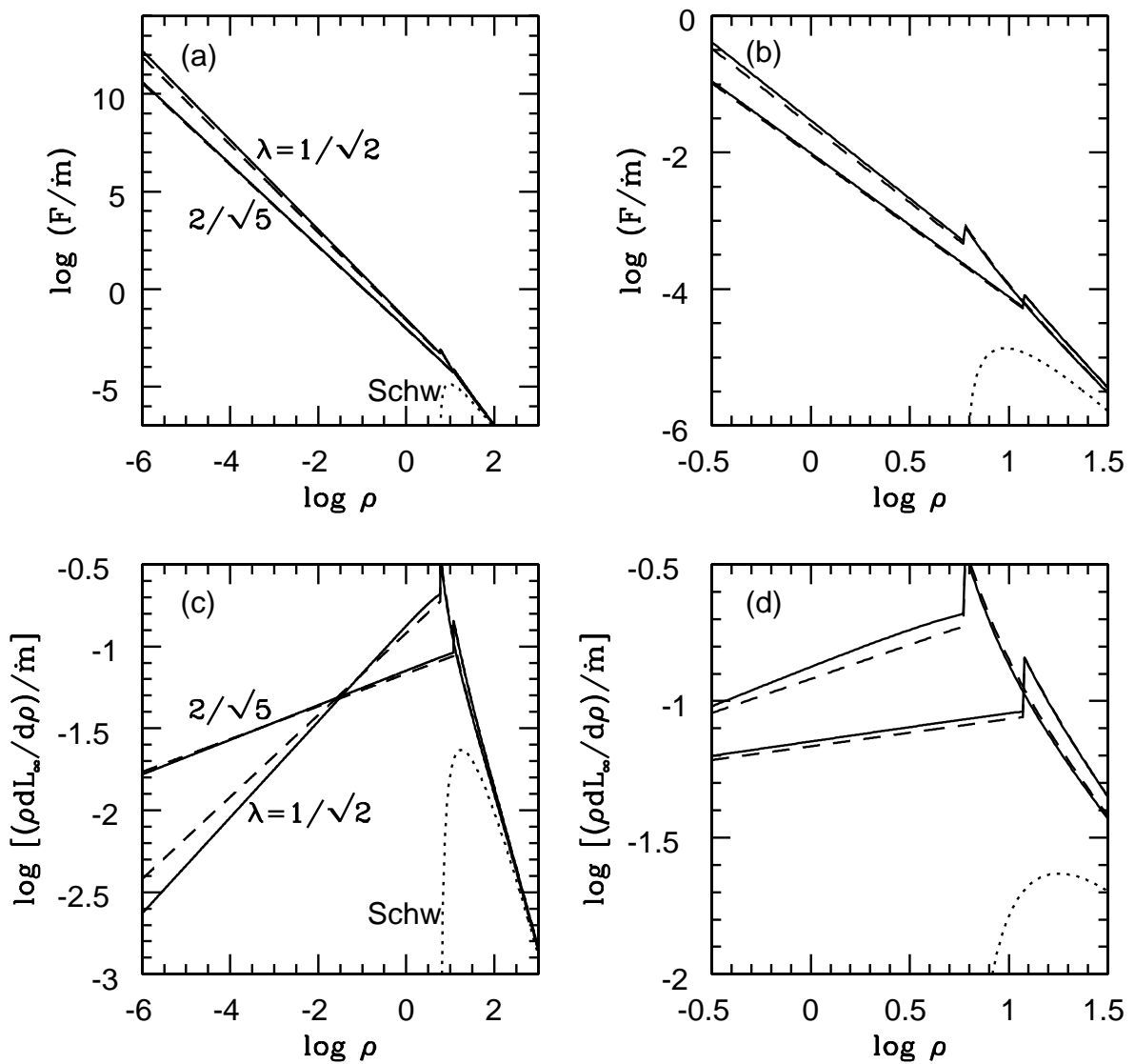}
\caption{(a) The two solid lines show the variation of the radiative
  flux $\mathcal{F}$ of an accretion disk, as measured by a local
  observer comoving with the fluid, for two perfect fluid singular
  models with $\lambda=1/\sqrt{2}$ ($\rho_b=6$) and
  $\lambda=2/\sqrt{5}$ ($\rho_b=12$). The dotted line corresponds to a
  disk around a Schwarzschild black hole, while the two dashed lines
  are for accretion disks in models with purely tangential pressure
  \cite{JMN}, which are discussed in section \ref{tangential}. (b)
  Close-up of the flux profiles near the matching radius. Note the
  discontinuity in the flux, which is caused by the discontinuous
  behavior of $\omega_{,\rho}$ in equation (\ref{flux}). (c) Profile
  of the differential luminosity reaching an observer at infinity,
  $d{\cal L}_\infty /d\ln\rho$, for the same models.  (d) Close-up of
  the region near the matching radius.}
\label{fig4}
\end{figure}

The solid curves in Figure \ref{fig4}(a) show the variation of
${\mathcal F}(\rho)$ vs $\rho$ for the two chosen models. The flux
diverges steeply as the gas approaches the center: ${\mathcal F} \sim
\rho^{-(3-\lambda)}$. This is not surprising, considering that the
cloud is singular in this limit. Perhaps more surprising is the
discontinuity in the flux at the boundary between the cloud and the
external vacuum metric, as seen clearly in Figure \ref{fig4}(b). While
all the quantities $E$, $L$, $\omega$, $g$ which are present in
equation (\ref{flux}) are continuous across the boundary, the
derivative $\omega_{,\rho}$ is not. The discontinuity in
$\omega_{,\rho}$ causes the jump in $F$ as $\rho$ crosses
$\rho_b$. The dotted lines in the two panels correspond to an
accretion disk around a Schwarzschild black hole. In this case, the
inner edge of the disk is at $\rho_{\rm in} = \rho_{\rm ISCO} = 6$,
and the flux cuts off at this radius.

We note of course that the flux ${\mathcal F}$ is a local 
quantity measured in the frame of
the fluid and is not directly observable.  A more useful quantity is
the luminosity ${\cal L}_\infty$ (energy per unit time) that reaches
an observer at infinity. The differential of ${\cal L}_\infty$ with
respect to the radius $\rho$ can be computed from $\mathcal{F}$ by the
following relation \cite{nt73}:
\begin{equation}
\frac{d{\cal L}_\infty}{d\ln\rho} = 4\pi \rho \sqrt{-g} E {\mathcal F}.
\label{dLdrho}
\end{equation}
Panels (c) and (d) in Figure \ref{fig4} show $d{\cal
  L}_\infty/d\ln\rho$ for the two models under consideration. We see
that the luminosity behaves in a perfectly convergent fashion as
$\rho\to 0$: $d{\cal L}_\infty /d\ln\rho \sim \rho^{1-\lambda}$. By
integrating this quantity over $\ln\rho$, we can calculate the net
luminosity ${\cal L}_\infty$ observed at infinity.  We have confirmed
that this is equal to $\dot{m}c^2$ for the two singular models. That
is the disk has 100\% efficiency --- it converts the entire rest mass
energy of the accreting gas into radiation.  The dotted lines in the
panels indicate the very different behavior of a disk around a
Schwarzschild black hole. Since such a disk is truncated at the ISCO,
the luminosity is much less. In this case, we recover the standard
result, ${\cal L}_\infty = (1-E_{\rm ISCO})\,\dot{m}c^2
=0.05719\,\dot{m}c^2$, i.e., an efficiency of 5.719\%.

Is it possible to distinguish observationally whether a given
accretion system is a black hole or one of the toy singular objects
described in this paper? We have seen above that the accretion
efficiencies are very different. However, the efficiency is not easily
determined via observations since there is no way to obtain an
indepedent estimate of the mass accretion rate $\dot{m}$.  A more
promising avenue is the spectral energy distribution of the disk
radiation.

Following standard practice, we assume that each local patch of the
disk radiates as a blackbody. Defining a characteristic temperature
$T_*$ as follows (where we have included physical units),
\begin{equation}
\sigma T_*^4 \equiv \frac{\dot{m}c^2}{4\pi (GM_T/c^2)^2}, \label{T*}
\end{equation}
where $\sigma$ is the Stefan-Boltzmann constant, the local blackbody
temperature of the radiation emitted at any radius $\rho$ with flux
$\mathcal{F}$ (eq.~\ref{flux}) is given by
\begin{equation}
T_{\rm BB}(\rho) = [{\mathcal F}(\rho)]^{1/4} T_*. \label{TBB}
\end{equation}
This radiation is transformed by gravitational and Doppler redshifts
by the time it reaches an observer at infinity and hence appears to
have a different temperature. The transformation depends on the
orientation of the observer with respect to the disk axis. To avoid
getting into too much detail, we simply use a single characteristic
redshift $z$, corresponding to an observer along the disk axis,
\begin{equation}
1+z(\rho) = [-(g_{tt}+\omega^2 g_{\phi\phi})]^{-1/2},
\end{equation}
and assume that the radiation emitted at radius $\rho$ has a temperature
at infinity, independent of direction, given by
\begin{equation}
T_\infty(\rho) = T_{\rm BB}(\rho)/(1+z).
\end{equation}
It is then straightforward to convert the differential luminosity
calculated in equation (\ref{dLdrho}) into the spectral luminosity
distribution ${\cal L}_{\nu,\infty}$ observed at infinity. The result
is
\begin{equation}
\nu {\cal L}_{\nu,\infty} = \frac{15}{\pi^4} \int_{\rho_{\rm in}}
^\infty \left(\frac{d{\cal L}_\infty}{d\ln\rho}\right)
\frac{(1+z)^4 (h\nu/kT_*)^4/\mathcal{F}}
{\exp[(1+z)h\nu/kT_*{\mathcal F}^{1/4}]-1}\, d\ln\rho.
\end{equation}
As a check, we have verified numerically that the integral over
frequency of the spectral luminosity $L_{\nu,\infty}$ obtained via the
above relation is equal to $\dot{m}c^2$ (100\% efficiency) for the
singular models and equal to $0.05719 \dot{m}c^2$ for the
Schwarzschild case (5.719\% efficiency).

\begin{figure}
\includegraphics[scale=0.6]{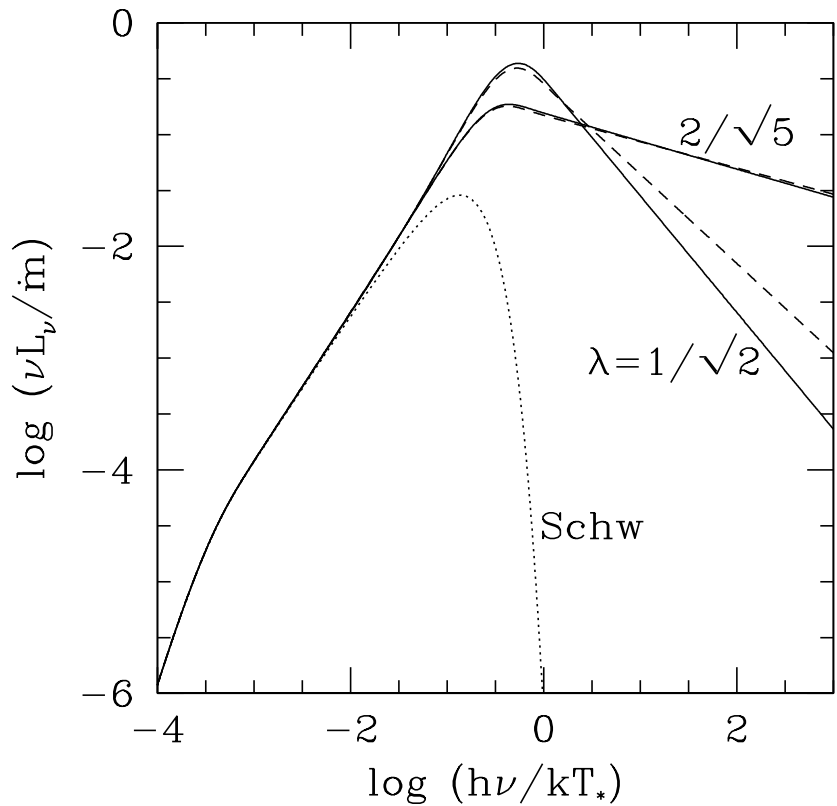}
\caption{Spectral luminosity distribution of radiation from the
  accretion disk models in Figure \ref{fig4}. The dotted line
  corresponds to a disk around a Schwarzschild black hole, the solid
  lines correspond to disks around two perfect fluid naked singularity
  models discussed in this paper, and the dashed lines are for disks
  around two tangential pressure naked singularity models discussed in
  \cite{JMN}.}
\label{fig5}
\end{figure}

Figure \ref{fig5} shows spectra corresponding to the same models
considered in Figure \ref{fig4}. For a given accretion rate $\dot{m}$,
the Schwarzschild black hole model gives a much lower luminosity than the
naked singular models. However, as we discussed earlier, this is not
observationally testable. More interesting is the fact that the
spectra have noticeably different shapes.  At low frequencies, all the
models have the same spectral shape $\nu L_\nu \sim \nu^{4/3}$, which
is the standard result for disk emission from large non-relativistic
radii \cite{FKR}. However, there are dramatic differences at high
frequencies.

The emission from a disk around a Schwarzschild black hole cuts off at
the ISCO radius $\rho_{\rm ISCO}=6$. Correspondingly, there is a
certain maximum temperature for the emitted radiation, which causes
the spectrum to cutoff abruptly at a frequency $h\nu/kT_*$ somewhat
below unity. The two perfect fluid singular models, on the other hand,
behave very differently. In these models, the disk extends all the way
down to $\rho\to0$. Consequently, the temperature $T_{\rm BB}$ of the
emitted radiation rises steadily and diverges as $\rho\to0$.  The
emission from all the inner radii combines to produce a power-law
spectrum at high frequencies 
\footnote{Note that models with
  $\lambda<1/3$ do not have a high energy power-law tail. These models
  belong to the class of models with $\lambda<1/\sqrt{2}$ which we
  have decided to ignore in this paper since their disks have a gap
  between the ISCO of the external Schwarzschild spacetime and the
  surface of the cloud.}: $\nu{\cal L}_{\nu,\infty} \sim
\nu^{-8(1-\lambda)/(6\lambda-2)}$.

As Figure \ref{fig5} shows, the high-frequency power-law segment of
the spectrum carries a substantial fraction (more than half) of the
total emission from disks around our naked singularity models. The
presence of this strong power-law spectrum is thus a characteristic
feature of these models which can be used to distinguish them
qualitatively from disks around black holes. Several well-known
astrophysical black hole candidates are known to have spectra with a
strong thermal cutoff, similar to that seen in the dotted line in
Figure \ref{fig5} \cite{thermal}. These systems show very little
power-law emission at high frequencies.  In these cases at least we
can state with some confidence that the central mass does not have a
naked singularity of the sort discussed in this paper.

\subsection{Comparison of the perfect fluid and tangential pressure models}
\label{tangential}
In our previous paper \cite{JMN}, we considered a relatively
restricted model in which the fluid in the cloud has non-zero pressure
only in the tangential direction. Because of vanishing radial
pressure, matching pressure across the boundary radius $\rho_b$
between the cloud and the external Schwarzschild spacetime was
trivial, and hence the solution was more easily determined. For radii
inside the cloud, the metric of the pure tangential pressure model
takes the form,
\begin{equation}
ds^2 = -H(\rho) dt^2+ \frac{d\rho^2}{(1-M_0)}
    +\rho^2 d\Omega
^2, \quad \rho \le \rho_b, \label{tangmetric} 
\end{equation}
with $H(\rho) = (1-M_0) \left(\rho/\rho_b\right)^{M_0/(1-M_0)}$ and $M_0 = 2M_T/\rho_b$, where $M_T$ is the total gravitational mass of the cloud as measured in
the vacuum exterior.

The close similarity of the tangential pressure model and the perfect
fluid model described in the present paper is obvious, e.g., compare
the above expressions with equations (\ref{metric-sol}) and
(\ref{MT2}). The constant $M_0$ in the tangential pressure model plays
the role of $\lambda$ for the perfect fluid case, and in both cases
this parameter determines the compactness of the cloud as measured by
its dimensionless radius $\rho_b/M_T$. The limits $M_0\to0$ and
$\lambda\to1$ in the two models correspond to infinitely large and
dilute clouds that are fully non-relativistic: $\rho_b/M_T \to
\infty$.  In this limit, the two models are essentially identical. One
minor difference between the models occurs in the opposite limit. For
the perfect fluid cloud, the most compact configuration we find has
$\lambda=0$, which corresponds to $\rho_b=4M_T$. In contrast, for the
tangential pressure model, the most compact physically valid
configuration has $M_0 = 2/3$, which corresponds to $\rho_b=3M_T$,
i.e., a more compact object.

Another difference between the two models is that the tangential
pressure cloud has a particularly straightforward metric in which
$g_{tt}=H(\rho)$ varies with radius as a simple power-law. Therefore,
all quantities behave as power-laws and the analysis is easy.  In the
case of the perfect fluid model, the extra matching condition on the
pressure at $\rho=\rho_b$ results in the metric coefficient $g_{tt}$
involving two power-laws. The term involving the coefficient $A$ in
equation~(\ref{metric-sol}) dominates as $\rho\to0$ and behaves just
like the lone power-law term in the tangential pressure cloud
case. The second term involving $B$ is required in order to satisfy
the pressure boundary condition and plays a role only as $\rho$
approaches $\rho_b$. This term causes various quantities like $E$,
$L$, $\omega$, etc. to deviate from perfect power-law behavior as
$\rho\to\rho_b$ (see Figs.~1--4).

As far as observables are concerned the two models behave quite
similarly. In Figs.~4 and 5, the two perfect fluid models with
$\lambda=1/\sqrt{2}$ and $2/\sqrt{5}$ have boundaries at $\rho_b=6$
and $12$, respectively, and their properties are shown by the solid
lines. For comparison, the dashed lines show results for two
tangential pressure models with $M_0=1/3$ and $1/6$ which have
boundaries at the same radii, $\rho_b=6$, $12$.  While the agreement
between the two sets of models is not perfect, as is to be expected
since the models are different, we see very good qualitative
agreement. In particular, note that, just as in the perfect fluid
models, the tangential pressure models too produce a strong power-law
high energy tail in the spectra of their accretion disks (Fig.~5),
which may be used to distinguish these models from disks around a
Schwarzschild black hole (dotted line).

\subsection{Comparison with the Newtonian singular isothermal sphere}
A simple and commonly used model in astrophysics is the singular
isothermal sphere. This is a spherically symmetric self-gravitating
object with an equation of state $p=\epsilon c_s^2$, where $p$ is the
pressure, $\epsilon$ is the density (this is usually called $\rho$ but
we use $\epsilon$ here since we have already defined $\rho$ to be the
radius), and $c_s$ is the isothermal sound speed. The model satisfies
the condition of hydrostatic equilibrium under the action of Newtonian
gravity. The solution is
\begin{equation}
\epsilon(\rho)=\frac{c_s^2}{2\pi G\rho^2}, \qquad
p(\rho) = \frac{c_s^4}{2\pi G \rho^2}, \qquad
M(\rho) = \frac{2c_s^2 \rho}{G}, \label {sis}
\end{equation}
where $M(\rho)$ is the mass interior to radius $\rho$. The variations of
$\epsilon$, $p$ and $M$ with $\rho$ are very reminiscent of the
variation of $\epsilon$, $p$ and $F$ in the perfect fluid relativistic
cloud model described earlier.

One deficiency of the basic singular isothermal sphere model is that
it extends to infinite radius, where the mass is infinite. In order to
obtain an object with a finite radius, one needs to change the
equation of state such that the pressure goes to zero at a finite
density. This is easily arranged as follows:
\begin{equation}
\epsilon(\rho)=\frac{c_s^2}{2\pi G\rho^2} = 
\frac{\rho_b^2}{\rho^2}\epsilon_b, \qquad
\epsilon_b \equiv \frac{c_s^2}{2\pi G\rho_b^2}, \qquad
p(\rho) = c_s^2 (\epsilon-\epsilon_b), \qquad
M(\rho) = 4\pi\epsilon_b r_b^2 \rho, \qquad \rho \le \rho_b, \label {sis2}
\end{equation}
where $\rho_b$ is the radius of the boundary and $\epsilon_b$ is the density
at that radius.

Circular orbits inside a singular isothermal sphere behave very
simply, with velocity and angular momentum given by
\begin{equation}
v_{\rm circ} = \sqrt{c_s}\, \rho = {\rm const}, \qquad
L = \sqrt{2}\, c_s \rho, \qquad \rho \le \rho_b.
\end{equation}
The constancy of $v_{\rm circ}$ is the chief attraction of the
singular isothermal sphere. It provides a simple way of reproducing
the observed flat rotation curves of galaxies.  The scaling of $L$
with $\rho$ in the singular isothermal sphere is exactly the same as the
variation of $L$ with $\rho$ in the perfect fluid relativistic
cloud. At first sight it appears that the flat rotation curve property
is not reproduced in the relativistic model. For example, $\omega \rho$
scales as $\rho^{1-\lambda}$. However, note that $\omega$ is defined
as $d\phi/dt$, where $t$ is the time measured at infinity. The appropriate
quantity to consider is $d\phi/dt_{\rm loc}$, where $t_{\rm loc}$ is
time measured by a local ZAMO (Zero Angular Momentum Observer) 
at radius $\rho$. The two times are
related by a factor of $[H(\rho)]^{1/2}$ which varies as $A
\rho^{1-\lambda}$ in the limit $\rho\to 0$. Thus, for ZAMOs, the rotation
velocity is indeed independent of radius in the deep interior of the
cloud. (Close to the boundary, there are small deviations because of
the term involving $B$ in $H(\rho)$).

Consider next the energy of circular orbits in the singular isothermal
sphere model. If we include the rest mass energy and add to it the
orbital kinetic energy and the potential energy, then a Newtonian
calculation gives for the energy of a particle of unit mass
\begin{equation}
E_{\rm circ} = 1-\frac{c_s^2}{c^2}\left(1+2\ln\frac{\rho_b}{\rho}\right),
\qquad \rho \le \rho_b.
\end{equation}
The weak logarithmic divergence at small radii is rarely a problem
since we generally have $c_s\ll c$. Nevertheless, the presence of the
logarithm implies that, in principle, at a sufficiently small radius,
the Newtonian model predicts a {\it negative} total energy for the
particle.  This is clearly unphysical.

It is reassuring that the logarithmic divergence is not present in our
relativistic cloud models.  Both the pefect fluid model and the
tangential pressure model have $E$ varying as a power-law with radius:
$E\sim \rho^{1-\lambda}$ for the perfect fluid case and $E\sim \rho
^{M_0/2(1-M_0)}$ for the tangential pressure case. In both cases, $E$
asymptotes precisely to zero as $\rho\to 0$ and does not go negative
anywhere. Effectively, the relativistic models, being more
self-consistent, regularize the logarithm of the Newtonian model by
replacing it with a power-law. The index of the power-law is nearly 0
in the Newtonian limit but becomes as large as unity for the maximally
compact configuration, viz., $\lambda=1$ and $M_0=2/3$ for the perfect
fluid and tangential pressure models, respectively.

\subsection{Other models}\label{other}
As we have seen above, the perfect fluid relativistic model described
in this paper is equivalent to the Newtonian singular isothermal
sphere model used in astrophysics.  The relativistic generalization
was obtained by assuming that the energy density $\epsilon(\rho)$ in
the relativistic model has the same functional form as $\epsilon(\rho)$ in
the Newtonian model (compare eqs.~\ref{epsilon2} and \ref{sis2}), and
then solving for the pressure $p$ and the spacetime metric. This
procedure can be followed with any other trial model of
$\epsilon(\rho)$ of interest.

One simple example is to consider a superposition of the toy perfect
fluid model described in this paper with the well known constant
density Schwarzschild interior.  The density profile then takes the
form
\begin{equation}
    \epsilon=\frac{M_0}{\rho^2}+M_1,
\end{equation}
which corresponds to 
\begin{equation}
F(\rho)=M_0\rho+M_1\rho^3/3.
\end{equation}  
The density in this model approaches the singular interior of our
perfect fluid model as $\rho\to 0$ but resembles a constant density
interior with $\epsilon=M_1$ as $\rho\to \rho_b$. It turns out that
the TOV equation can be explicitly integrated in this case, though the
expression for $p$ is fairly complicated and involves hypergeometric
functions.

Other examples of more interest to astrophysics could be similarly
considered. One natural generalization of the singular isothermal sphere is the
Jaffe density profile \cite{Jaffe},
\begin{equation}
    \epsilon=\frac{M_0r_0}{\rho^2(\rho_0+\rho)^2},
\end{equation}
where $M_0$ is a constant and $\rho_0$ describes the characteristic
radius of the object. This density profile corresponds to
\begin{equation}
    F(\rho)=\frac{M_0\rho}{(\rho_0+\rho)}.
\end{equation}
The Jaffe model behaves just like the singular isothermal sphere as
$\rho\to0$, yet it has a finite total mass given by $M_0$ and hence does
not need to be artifically truncated as we had to in the case of the
singular isothermal sphere.  The Jaffe model is a special case of a
more general class of models, the Dehnen density profile
\cite{Dehnen},
\begin{equation}
\epsilon=\frac{(3-\gamma)M_0\rho_0}{\rho^\gamma(\rho+\rho_0)^{4-\gamma}}.
\end{equation}
The Jaffe profile corresponds to $\gamma=2$, while the case with
$\gamma=1$ is known as the Hernquist model \cite{hernquist}. The
Dehnen profile implies a mass function
\begin{equation}
    F(\rho)=M_0\left(\frac{\rho}{\rho_0+\rho}\right)^{3-\gamma}.
\end{equation}
Finally, we could also consider the Navarro-Frenk-White (NFW) profile
\cite{NFW}, which is given by
\begin{equation}
    \epsilon=\frac{M_0}{\rho(\rho_0+\rho)^2},
\end{equation}
where again $M_0$ is a constant. This model has a logarithmically
diverging mass as $\rho\to\infty$, and is thus a little less
attractive.

It is not easy to solve the TOV equations for the pressure $p$ and the
spacetime metric of any of the above models analytically. However,
numerical solutions are easily obtained. Note that all these popular
models technically have naked singularities at their
centers. Obtaining relativistic generalizations along the lines
followed in this paper would be worthwhile.

\section{Concluding remarks}\label{conclusion}
In the present paper we considered a self-bound spherical cloud
of perfect fluid and derived static non-vacuum solutions of the
Einstein equations which posseses a central naked singularity. We
showed that these solutions could be obtained asymptotically as the
final result of the slow collapse of a massive matter cloud. The
solutions described here closely parallel those we obtained in
\cite{JMN} for a fluid with pure tangential pressure.

We studied the properties of steady thermal accretion disks in our
naked singularity spacetimes.  Focusing on those models that have a
disk extending continuously from a large radius $\rho$ down to
$\rho\to0$, i.e., models with $\lambda \geq 1/\sqrt{2}$ for the
perfect fluid case and $M_0\leq 1/3$ for the tangential pressure case,
we showed that accretion disk spectra would consist of a
multi-temperature blackbody at low frequencies joining smoothly to a
power-law at high frequencies.  Notably, the disk luminosity would be
dominated by the high-energy tail, which is a characteristic feature
of these models. The spectrum of an equivalent accretion disk around a
black hole would have only the low-frequency multi-temperature
blackbody component and would be missing the high frequency tail (or
at best have a weak tail).

Accretion disks around astrophysical black hole candidates do show
power-law tails, but these are usually interpreted as coronal emission
from hot gas above the disk. In those cases where the disk emission is
definitely thermal, e.g., in the so-called Thermal State of accreting
stellar-mass black holes \cite{rem_mccl}, the spectrum is invariably
dominated by the low-frequency multi-temperature emission and the
power-law tail tends to be quite weak. A number of stellar-mass black
holes have been observed in the Thermal State \cite{mccl13}. In the
case of these systems at least we may conclude that the central
compact objects are not naked singularities of the type discussed in
the present paper; the objects are presumably true black
holes. However, it is not possible to state anything with certainty in
the case of other black hole candidates that have not been observed in
the Thermal State.

The existence of horizons in stellar mass astrophysical 
sources has been a matter of discussion in recent years (see for example
\cite{horizonevidence}, \cite{horizon} and references therein).
Some observations suggest that the departure of such objects 
from black holes must be very small
\cite{Doeleman}.
All the same, when it comes to more massive 
sources, excess luminosity from Ultra Luminous X-ray Sources cannot at 
present be explained fully by the usual black hole accretion 
disk models and seem to require the existence of as yet undiscovered 
intermediate mass black holes
\cite{ULX}.

Since there is a strong thermal cutoff at high frequencies  
present in the spectrum of the black hole models while it is absent in 
the perfect fluid naked singularity models studied here, this  
strenghtens the argument that the astrophysical sources 
that exhibit the same behaviour would be black holes.
On the other hand, ultra luminous sources exist in the universe 
and at present it is not clear if it will be possible to fit all the 
observations within the black hole paradigm.
It is possible that different kinds of sources exist in nature,
besides black holes, stars and neutron stars. These
could be of the singular type discussed here or could be regular
objects composed of ordinary or exotic matter. All these could 
possibly form from collapse processes as described here and 
they would have their distinctive observational features 
(see for example 
\cite{new}).
In such a scenario, the comparison and models such as the ones 
discussed here may be useful to understand better and analyze
future observations.

For the toy models presented here we found that the radiant 
energy flux and the spectral energy distribution are much greater as 
compared to a black hole of the same mass, and therefore we obtained 
indications that if some sources of similar kind do exist in the 
universe it might be possible to distinguish them observationally 
from black holes.

We also found here that the qualitative features described in
\cite{JMN}
are preserved when we consider perfect fluid sources, instead of 
sources sustained only by tangential stresses. We can therefore 
conjecture that the increased flux and luminosity may be a generic 
feature of any source with a singularity at the center where the
accretion disk can in principle extend until $r=0$. 
As it is known, perfect fluids 
are important in the context of astrophysics as they can be 
used to model many sources and objects of astrophysical relevance.
Of course, different matter models will imply 
different luminosity spectrums that might or might not eventually 
be distinguished from one another.

Finally, we note that these models require a certain fine 
tuning since all the collapsing shells must have the right velocity 
so that the effective potential leads asymptotically to 
$\dot{R}=\ddot{R}=0$. Nevertheless the class of static final 
configurations can be understood in the sense of an idealized 
approximation for a slowly evolving cloud with small velocities. 
In this sense these scenarios appear to be generic as they 
approximate a wide variety of slowly evolving matter 
clouds.

\end{document}